\documentclass[11pt]{article}

\usepackage[utf8]{inputenc}
\usepackage{graphicx}
\usepackage[colorlinks,allcolors=red]{hyperref}
\usepackage{amsmath,amssymb,amsfonts,mathtools}

\usepackage{indentfirst}
\usepackage{lipsum}
\usepackage{wrapfig}
\usepackage{graphicx,overpic}
\usepackage{siunitx}
\usepackage{textcomp}
\usepackage{titlesec}
\usepackage{tikz}
\usepackage{bm}
\usepackage{tcolorbox}
\usepackage{xcolor}
\usepackage{subcaption}
\usepackage{graphicx}
\usepackage{pgfplots}
\graphicspath{{img/}} % includegraphics path
\usepackage{multirow}
\usepackage{hhline}

\usepackage{bm}
\newcommand{\vect}[1]{\boldsymbol{\mathbf{#1}}}
\newcommand{\bu}{\mathbf u}

\newcommand{\bF}{\mathbf F}

\newcommand*\diff{\mathop{}\!\mathrm{d}}

\newcommand{\gradG}{\nabla_{\Gamma}}
\newcommand{\laplG}{\Delta_{\Gamma}}

\newcommand{\divG}{{\mathop{\,\rm div}}_{\Gamma}}

\def\el {\nonumber }

\begin{document}

\title{On fusogenicity of  positively charged phased-separated lipid vesicles: experiments and
computational simulations}

\author{Y.~Wang$^{1,\dag}$, Y.~Palzhanov$^{2,\dag}$, D.~T.~Dang$^{1}$,  A.~Quaini$^{2,*}$, M.~Olshanskii$^{2}$, S.~Majd$^{1,*}$}

\maketitle

\begin{center}
\noindent $^{1}$Department of Biomedical Engineering, University of Houston, 3551 Cullen Blvd, Houston TX 77204\\
\texttt{ywang147@uh.edu; thiendang1197@gmail.com; smajd9@central.uh.edu}

\vskip .3cm
\noindent $^{2}$Department of Mathematics, University of Houston, 3551 Cullen Blvd, Houston TX 77204\\
\texttt{ypalzhanov@uh.edu; aquaini@uh.edu; maolshanskiy@uh.edu}
\end{center}

\vskip .3cm
\noindent $^{\dag}$ Equal contribution \\
$^{*}$ Corresponding authors

\vskip .3cm
\noindent{\bf Abstract} 
This paper studies the fusogenicity of cationic liposomes in relation to their surface distribution of cationic lipids and utilizes membrane phase separation to control this surface distribution. It is found that concentrating the cationic lipids into small surface patches on liposomes, through phase-separation, can enhance liposome's fusogenicity. Further concentrating these lipids into smaller patches on the surface of liposomes led to an increased level of fusogenicity. These experimental findings are supported by numerical simulations using a mathematical model for phase-separated charged liposomes. Findings of this study may be used for design and development of highly fusogenic liposomes with minimal level of toxicity. 


\section{Introduction}

\indent Nano-scale liposomes have proven to be highly effective and versatile drug delivery vehicles 
as they rely on two mechanisms for cellular uptake: endocytosis \cite{miller1998liposome,takikawa2020intracellular,manzanaresendocytosis} 
and membrane fusion \cite{duzgunecs1999mechanisms,yang2016drug,hui1996role}. 
Membrane fusion, that entails integration of two different membranes, is particularly appealing for delivery 
of macromolecules because through this mechanism liposomes deliver their encapsulated cargo directly into the cytoplasm. 
Liposomes that contain cationic lipids \cite{hui1996role,almofti2003cationic}, 
such as 1,2-dioleoyl-3-trimethylammonium-propane (DOTAP), are known 
for their high fusogenicity \cite{simberg2004dotap}. Cationic lipids, with their conical shape and cationic headgroup, are critical 
for fusion \cite{liu2020barriers,hoffmann2020complex,kolavsinac2019influence}. 
While these lipids are typically non-toxic at lower concentrations, 
concerns arise regarding their toxicity when used at higher concentrations, attributed to their 
tetrasubstituted ammonium moiety \cite{lv2006toxicity}. Therefore, designing delivery liposomes that offer both 
high fusogenicity and low toxicity is a challenge. This challenge may be overcome by controlling 
the surface density of cationic DOTAP on the surface of liposomes using the membrane phase separation phenomenon. 

Phase separation is a fundamental process that occurs in multicomponent lipid membranes 
with substantial unfavorable interactions among their lipid components \cite{heberle2011phase}. In such membranes, 
segregation of lipids based on their favorable interactions leads to phase separation. This phenomenon 
regulates molecular organization in membranes and thus can be used to control the surface density of the 
membrane's components. Ternary mixture of DOPC:DPPC:Chol is an example of a phase-separating 
composition that can, for instance, form a tightly-packed liquid ordered ($L_o$) phase and a loosely-packed 
liquid disordered ($L_d$) phase at certain molar ratios. We previously combined experiments and modeling to 
investigate the phase behavior in this lipid mixture \cite{zhiliakov2021experimental,WANG2022183898}. 
Here, we aim to explore the use of phase-separation 
in DOTAP:DOPC:DPPC:Chol mixture %\anna{(Maybe this is obvious, but shouldn't we say that we're replacing a percentage of DOPC with DOTAP?)} 
to modulate surface density of DOTAP on liposomes and hence their fusogenicity.    

\begin{figure}[htb]
	\centering
	\includegraphics[width = .6\textwidth]{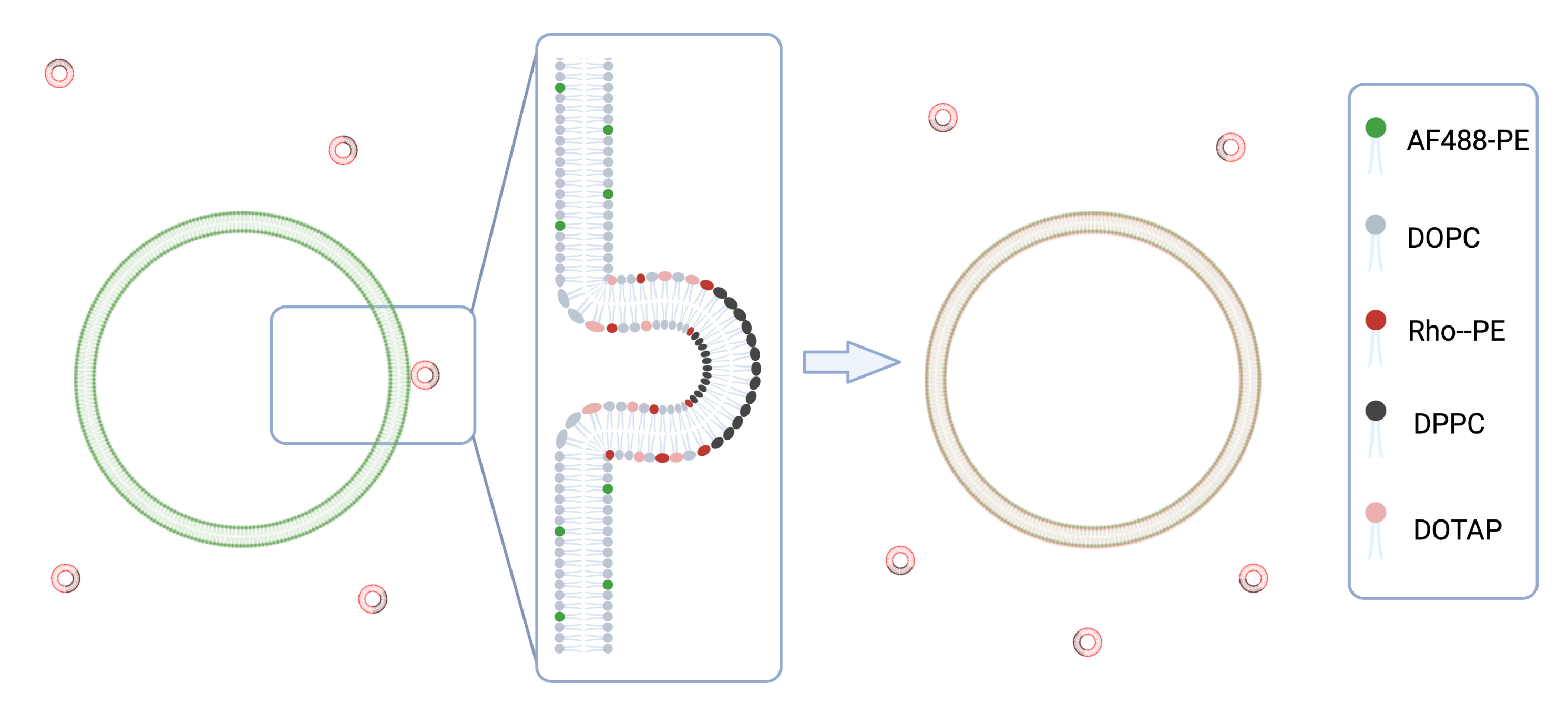}
	\caption{Schematic illustration of the phase-separated cationic SUVs (labeled with red-fluorescent lipids) fusing in to GUVs (labeled with green-fluorescent lipids).}
	\label{fig1}
\end{figure}

We hypothesize that concentrating DOTAP into small patches on the liposome's surface, through phase separation, 
can enhance the liposome's fusogenicity without the need for high DOTAP concentrations. We further postulate 
that liposomes with the smallest patch area (i.e., the highest local density of DOTAP when the amount of DOTAP
is kept fixed) would exhibit the highest 
level of fusogenicity into target membranes when compared to other liposomes with similar DOTAP content. 
To test these hypotheses, we examine the fusogenicity of nano-scale liposomes (referred to as small unilamellar vesicles - SUVs) 
of three different phase-separating compositions containing DOTAP (referred to as patchy liposomes - PAT) into micron-sized liposomes 
(referred to as giant unilamellar vesicles - GUVs) as model target membranes. Fluorescence microscopy was used as a tool to assess the level of SUV fusogenicity. To enable fluorescence microscopy as a gauge tool, different fluorescent lipids, Rho-PE and AF488-PE, were added to SUV and GUV membranes, respectively. The setup is schematically illustrated in Fig.~\ref{fig1}.

To complement our experimental studies, we further advanced the computational platform developed for 
\cite{zhiliakov2021experimental,WANG2022183898}. Specifically, we included the electrostatic interaction
between SUVs and GUVs into the surface Navier--Stokes--Cahn--Hilliard (NSCH) phase-field model, which accounts
for phase behavior, domain formation, and membrane fluidity in ternary membrane compositions. In \cite{WANG2022183898},
we validated the continuum-based NSCH model without electrostatic interaction against experimental data and showed
that our model predicts membrane phase behavior in a reliable and quantitative manner in the absence of DOTAP.
The extension of the model presented in this paper enables predictions when DOTAP is present.
This is a crucial step towards efficient computer-aided design of liposomes that
use cationic lipids for high fusogenicity and membrane phase-separation for limited toxicity.

\section{Materials and Methods}

\subsection{Experimental Approach}

\subsubsection{Materials}

Lipids 1,2-dioleoyl-sn-glycero-3-phosphocholine (DOPC), 1,2-dipalmitoyl-sn-glycero-3-phospho\linebreak
choline (DPPC), 1,2-dioleoyl-3-trimethylammonium-propane (DOTAP), 1,2-dipalmitoyl-sn-glycero-3-phosphoethanolamine-N- (lissamine rhodamine B sulfonyl) (Rho-PE) and 1,2-dioleoyl-sn-glycero-3-phosphoethanolamine-N-(TopFluor\textregistered AF488) (AF488-PE) were from Avanti Polar Lipids (Alabaster, AL). We purchased the sucrose from Avantor (Radnor Township, PA). Cholesterol was purchased from Sigma Aldrich (Saint Louis, MO) and chloroform from Omnipure (Caldwell, Idaho). All lipid stock solutions were prepared in chloroform. Indium tin oxide (ITO) coated glasses and microscope glass slides were from Thermo Fisher Scientific (Waltham, MA) and coverslips were bought from Corning (Corning, NY). ITO plates were cleaned with chloroform, ethanol and DI water before use. Microscope slides and coverslips were cleaned with ethanol and DI water.

\subsubsection{Preparation of GUVs}

We employed a modified version of electroformation \cite{kang2013simple,zhiliakov2021experimental}
to form GUVs. To this end, \SI{35}{\micro\litre} of an aqueous dispersion of SUVs was 
deposited as small droplets onto two ITO plates. The droplets were left overnight to dry. 
Subsequently, a thin PDMS frame with integrated tubing was assembled between the ITO plates to create a chamber. 
To rehydrate the dried lipids, a solution of \SI{235}{\milli\mole} %235 mM 
sucrose was slowly injected into the chamber. The device was then placed in a 60\textdegree{}C oven to exceed the highest lipid melting temperature in the mixture (in this case, DPPC with a melting temperature of 41.2\textdegree{}C) for phase separated liposome or room temperature for DOPC GUVs.

To induce vesicle formation, an AC electrical field was applied using a function waveform generator (4055, BK Precision, Yorba Linda, CA). 
The frequency was set at \SI{50}{\hertz}, and the electric field was gradually increased to 2 Vpp at a rate of \SI{100}{\milli\volt\per\minute}.
%100 mV/min. 
This field was kept for 3 hours. Once the vesicles were formed, the frequency was reduced to \SI{1}{\hertz} for 30 minutes for GUVs detachment.

%\anna{This section is titled ``Preparation of Giant Unilamellar Vesicles''. Since we talk about the SUVs too, should we change the title 
%of the section or move the paragraph below to the next subsection?}
The SUVs used in the electroformation process were prepared using dehydration-rehydration followed by tip sonication. 
In short, in a \SI{5}{\milli\litre} pearl-shaped flask, DOPC and fluorescent lipid, AF488-PE, was mixed with chloroform. 
The details of this portion of procedure are described in Sec.~\ref{sec:prep}. 
Next, the size of the SUVs was reduced through tip sonication using a 55-Watt Sonicator Q55 (Qsonica, Newtown, CT). 
The sonication process involved 30 seconds of resting followed by 1 minute of sonication at \SI{10}{\hertz}. This sonication-rest 
cycle was repeated 20 times to obtain a clear solution of SUVs. 

\subsubsection{Preparation and Characterization of SUVs for Fusion Experiments}\label{sec:prep}

SUVs used for fusion experiments were formed by dehydration-rehydration followed by extrusion, 
to provide a narrower size distribution. %These nanoliposomes will be referred to as small unilamellar vesicles (SUVs)
% in the following sections. 
 In brief, a solution of the desired lipid mixture including the fluorescent-lipid, Rho-PE, 
 in chloroform was prepared. Solution was placed in \SI{10}{\milli\litre} round flask and dried under vacuum using 
 a rotary evaporator (Hei-Vap, Heidolph, Germany) for 2 hours. The produced thin lipid film was then hydrated 
 using diluted (\SI{235}{\milli\mole}) PBS at a final lipid concentration of \SI{3}{\milli\mole}, % 3 mM, 
 and then extruded through polycarbonate membranes with \SI{100}{\nano\meter} pores (Cytiva, Marlborough, MA) for 25 times. 
Both rehydration and extrusion were done at room temperature for homogenous SUVs and at 60\textdegree{}C 
 for phase-separating SUVs. 
Fusion experiments applied both homogenous and phase-separating SUVs. For the phase-separating SUVs, lipid composition 
DOPC:DPPC:Chol with three different molar ratios were selected (see Table \ref{tab:comp}), 
in which DOPC was partially replaced with DOTAP.
We included Rho-PE to enable fluorescence microscopy.

\begin{table}[htb]
\begin{center}
 \begin{tabular}{ | c |  c |  c |  c |  c | }
\hline
Composition  & DOPC & DPPC & Chol & Rho-PE \\
\hline
Homo & 99.4\% & 0\% & 0\% &0.6\% \\
\hline
PAT1 & 59.4\% & 20\% & 20\% &0.6\% \\
\hline
PAT2 & 41.9\% & 42.5\% & 15\% &0.6\%\\
\hline
PAT3 & 24.4\% & 50\% & 25\% &0.6\%\\
\hline
\end{tabular}
\caption{Lipid composition for the examined liposomes.
}\label{tab:comp}
\end{center}
\end{table}

A Malvern Zetasizer machine (Nano-ZS, Malvern Instruments, Malvern, UK) was used to characterize 
the SUVs for size distribution (via dynamic light scattering) and zeta potential (via laser Doppler electrophoresis).

\subsubsection{Imaging and Analysis}

For fusion experiments, SUVs and GUVs were mixed at 1:1 molar ratio in microtubes and incubated at 37\textdegree{}C
for \SI{10}{\minute}. The sample was then collected and placed on a clean microscope glass slide. Double-sided tape was 
used between the glass slide and a coverslip to create a chamber for imaging. All the images were acquired using Zeiss 
LSM 800 confocal laser scanning microscope (Zeiss, Germany). Confocal images were obtained using 63$\times$ oil objective 
with NA of 1.40 using \SI{561}{\nano\meter} and \SI{488}{\nano\meter} wavelength lasers. Over 20 images from different 
areas of each sample were captured at each time point and a minimum of 25 GUVs per sample were used for analysis. 

The analysis was done using Zen 3.4 software (Zeiss, Germany). The green channel images were used to determine 
the location of the vesicles. Two circles $V_o$ and $V_i$ were drawn at the outer and inner borders of each vesicle, 
to isolate the signal from the membrane. Similarly-sized circles $B_o$ and $B_i$ were used to measure the 
background fluorescence intensity. The mean fluorescence intensity ($I$) and the area ($A$) of the isolated region 
were analyzed by the software. The fluorescence intensity was then calculated as:
\begin{align}
I_V = \frac{A_{V_o} I_{V_o} - A_{V_i} I_{V_i}}{A_{V_o} - A_{V_i}}, \quad 
I_B = \frac{A_{B_o} I_{B_o} - A_{B_i} I_{B_i}}{A_{B_o} - A_{B_i}}, \quad
I_M = I_V  - I_B, \el
\end{align}
where $I_V$ and $I_B$  represent the red fluorescence intensity of the vesicle membrane and background, respectively, 
and $I_M$ represents the background-subtracted fluorescence intensity of the GUV membrane. 
Only when $I_M$ for a GUV was determined to be larger than 1, the vesicle was considered as a GUV showing fusion. 
%\MO{After a quick reading about fluorescence intensity, my understanding that it is  measured here in RFU units. Shouldn't we say what is the reference measurement then or this is insignificant piece of information?} 
%\anna{Sheereen's answer: fluor signal measured in images has arbitrary units and I believe the provided description is sufficient.}

\subsection{Computational Approach}\label{sec:comp}

%\subsubsection{Mathematical model}\label{sec:math_model}

In order to reproduce and predict experimentally observed phenomena, the mathematical model needs to account
for three major physical factors: i) phase separation, ii) surface density flow, and 
iii) electrostatic forces.
The thermodynamically consistent NSCH model introduced in \cite{Palzhanov2021},
and validated against experimental data in \cite{WANG2022183898}, accounts only for i) and ii), i.e., only
phase separation and flow phenomena occurring in lipid membranes can be {modeled} computationally. 
In this paper, we extend the NSCH model to include the electrostatic forces between the positively charged lipids in the SUVs
and the GUVs, whose average measured zeta potential is negative (see Table \ref{tab:zeta}).

In order to state the model, let $\Gamma$ be a sphere representing an SUV with a \SI{120}{\nano\metre} 
%\anna{(Sheereen and Yifei, is that right?)} %\SI{10}{\micro\metre} 
diameter and let $c_i$ be a fraction of elementary surface area occupied by %\MO{lipids in} 
phase $i$, with $i = L_o, L_d$. %Let phase 1 be the $L_o$ phase. 
We choose $c = c_{L_o}$, $c\in [0,1]$, as the representative surface fraction. 
Let $\rho_{L_o}$ and $\rho_{L_d}$ be the densities and $\eta_{L_o}$ and $\eta_{L_d}$ the dynamic viscosities of the two phases.
Then, the density and viscosity of the mixture can be written as $\rho= \rho(c) = \rho_{L_o} c+ \rho_{L_d} (1-c)$
{and} $\eta=\eta(c)=\eta_{L_o} c+\eta_{L_d}(1-c)$, respectively. Let $\bu$ be the {area-}averaged
tangential velocity in the mixture, $p$ the thermodynamic interfacial pressure, and $\mu$ the chemical potential.
Finally, let $\bF_e$ denote the electrostatic force per unit surface {area} acting on the SUV. 
The NSCH system with electrostatic forcing that governs the evolution of $c$, $\bu$, $p$, and $\mu$ in time $t$ and space $\vect x \in\Gamma\subset\mathbb R^3$
is given by:
\begin{align}
&\small  \underbrace{\rho(\partial_t \bu + (\gradG \bu)\bu)}_{\text{inertia}} - \underbrace{\operatorname{\mathbf{div}}_{\Gamma}(2\eta E_s(\bu))+ \gradG p}_{\text{lateral stresses}} =  \bF_e \underbrace{-\sigma_\gamma \epsilon^2 \divG \left(  \gradG c \otimes\gradG c \right)}_{\text{line tension}} + \underbrace{{M \theta(\gradG(\theta\bu)\,)\gradG \mu}}_{\text{chemical momentum flux}} \label{grache-1m} \\
& \underbrace{\divG \bu  =0}_{\text{membrane inextensibility}}&  \label{gracke-2}\\
&\underbrace{\partial_t c +\divG(c\bu)}_{\text{transport of phases}}-  \underbrace{\divG \left(M \gradG \mu \right)}_{\scriptsize\begin{array}{c}\text{phase masses exchange}\\ \text{Fick's law}\end{array}}  = 0,\qquad
\mu = \underbrace{ f_0'(c) - \epsilon^2 \laplG c}_{\text{mixture free energy variation}}  \label{gracke-4}
\end{align}
on $\Gamma$ for $t \in (0,t^{\rm final}]$. In eq.~\eqref{grache-1m}-\eqref{gracke-4}, $\nabla_\Gamma$ stands for the tangential gradient, $\Delta_\Gamma$ for the Laplace--Beltrami operator,   
$E_s(\bu) = \frac12(\nabla_\Gamma \bu + (\nabla_\Gamma \bu)^T)$ is the Boussinesq--Scriven strain-rate tensor, and
$\divG$ is the surface divergence. 
Eq.~\eqref{gracke-4} provides the definition of the chemical potential, with $f_0(c) = \frac{1}{4}\,c^2\,(1 - c)^2$
being the double-well thermodynamic potential and parameter~$\epsilon > 0$ representing
the width of the (diffuse) interface between the phases. In addition, $\sigma_\gamma$ is line tension coefficient, 
$M$ is the mobility coefficient (see \cite{Landau_Lifshitz_1958}), and 
$\theta^2 = \frac{d\rho}{dc}$. 
Problem \eqref{grache-1m}-\eqref{gracke-4} 
models the total exchange of matter between phases (eq.~\eqref{gracke-4}) 
with surface flow described in terms of momentum conservation (eq.~\eqref{grache-1m}) 
and area preservation (eq.~\eqref{gracke-2}). 

To set viscosity and line tension, we referred to experimental work from \cite{SAKUMA20201576,Heftberger2015,C3SM51829A,KUZMIN20051120}. 
In \cite{WANG2022183898}, we calculated the value of density for each phase using the 
estimated molecular weight and molecular surface area for the corresponding phase. However, 
those values do not take into account the fact that the vesicle is loaded with and surrounded by an aqueous solution. 
Hence, in this paper we have increased the values to account for the 
``added mass'' coming from such solution.  
Table \ref{tab:physical_param} reports the domain ($L_o$ phase) area fraction $a_D$ and
the values or range of values for viscosity, line tension, and density 
for the compositions under consideration. 
%(see Table \ref{tab:comp} for more details  \MO{should it be a reference to Table~\ref{tab:physical_param} instead? Also, what is abbreviation PAT standing for? Maybe it makes sense to decrypt the acronym? }
%\anna{Maybe we can make a quick mention to the compositions in Sec.~2. In the previous paper, we had
%``''}).

\begin{table}[htb]
\begin{center}
 \begin{tabular}{ | c | c | c |  c |  c |  c |  c | }
\hline
Composition  &  ${a_D}$ & $\rho_{L_o} $ & $\rho_{L_d}$ & $\eta_{L_o}$ & $\eta_{L_d}$ & $\sigma_\gamma$   \\
\hline
PAT1 & $ 10.8\%$ (15\textdegree{}C) & 1401 & 1172 & $0.5-6$ & $0.2-0.4$ & $1.2-1.4$ \\
\hline
PAT2 & $ 34.57\%$ (17.5\textdegree{}C)  & $1401$ & $1172$ & $0.43-5.7$  & $0.2-0.4$ & $1.2-1.6$  \\
\hline
PAT3  & $70.37\%$ (15\textdegree{}C) & 1435 & 1172 & $5-8$ & $0.2-0.4$ & $1.2-1.8$  \\
\hline
\end{tabular}
\caption{
%\YP{For PAT2-3 we used numbers from previous paper, For PAT1 it is a composition from paper with Alex, there NS wasn't implemented yet and in that paper we don't have these parameters, so we use similar numbers to PAT2.}
%\anna{I thought we increased density, didn't we?}
Domain ($L_o$ phase) area fraction $a_D$ (at the given temperature), 
value or range of values for the density of liquid ordered ($\rho_{L_o}$) and liquid disordered ($\rho_{L_d}$) phases 
in Kg/(mol$\cdot$\r{A}$^2$), viscosity of liquid ordered ($\eta_{L_o}$) and liquid disordered ($\eta_{L_d}$) phases in {$10^{-8}$} Pa$\cdot$s$\cdot$m, and line tension in pN  for the three membrane compositions under consideration.} \label{tab:physical_param} 
\end{center}
\end{table}

Like in our previous works \cite{zhiliakov2021experimental,WANG2022183898}, 
we consider degenerate mobility $M = D c\,(1-c)$. Parameter $D$ is related to
thermodynamics  properties of matter, just like parameter  $\epsilon$ in eq.~\eqref{gracke-4}, which 
is the width of transition layer between ordered and disordered phases. Since the direct evaluation
of both $D$ and $\epsilon$ is not straightforward, in \cite{zhiliakov2021experimental} we relied on a 
data driven approach for their estimation. Our estimate for $D$ is $10^{-5}(\mbox{cm})^2$s$^{-1}$, %\anna{(Yerbol, are these numbers right?)} 
%depending on the membrane composition, 
while we found that $\epsilon=$\SI{1}{\nano\meter} %\anna{(for a 10 micron GUV, we used
%\SI{0.1}{\micro\meter} but since this is a SUV I'm changing the order of magnitude)}
 is a good estimation for $\epsilon$.

In the simulations, we exposed one SUV to one GUV. Because the GUVs are significantly larger than the SUVs, 
the curvature of a GUV is negligible {at the scale given by the size} of an SUV. Hence, we will approximate a GUV with a plane
for the computation of the electrostatic force $\textbf{F}_e$. Therefore, the electric field $\textbf{E}$ generated by a GUV
{can be (locally) computed by}:
\begin{equation}\label{eq:E}
\boldsymbol{E} = \frac{\sigma}{2\varepsilon_0}, 
\end{equation}
where $\sigma$ is the GUV surface charge density and $\varepsilon_0$ is the vacuum permittivity ($8.85\cdot10^{-12}$ \SI{}{\farad\per\meter}). 
The value of $\sigma$  is estimated from a linear approximation of
Grahame's formula \cite{interfacesbook}, which is valid in low-potential situations:
\begin{equation}
	\sigma   \approx \varepsilon\cdot \varepsilon_0 \cdot\kappa \cdot \Psi_0 , \quad  \Psi_0 = \frac{\zeta}{\exp(-\kappa\cdot x)},
	%= (8*R*T*c*\epsilon*\epsilon_0)^{1/2} \sinh(\frac{z_i*F}{2*R*T}*\Psi_0)
	\label{grahame}
\end{equation}
where $\varepsilon$ is the relative permittivity of water (about 80 at 20\textdegree{}C), %\anna{Yerbol, your notes reported
%a value of 70, while a quick google search gives 80, are you sure of the number in your notes?} \YP{we can change it to 80}),
%\anna{(Sheereen and Yifei: please advise if the value of water is not what we should consider)}
%\anna{of what medium?} \YP{this is apparoximate for water, I asked Yifei about this but I didn't get any comments. I kept water.}, 
$\kappa$ is the Debye length parameter for a  NaCl solution (10/7 $\text{nm}^{-1}$ \cite{chibowski2016zeta}), %\anna{Yerbol, please specify what solution this corresponds to because the length varies depending on the ionic strength}\YP{it is a NaCl solution from \cite{chibowski2016zeta}}),
%\anna{(this also depends on the medium, right?)} \YP{Yes, this depends but we take it from paper \cite{chibowski2016zeta}, units should be correct. If Yifei and Dr.Majd think that this params should be adjusted a little that's fine, because playing with these params little wouldn't change our current results and trends much, maybe a little.} , 
$\Psi_0$ is the surface potential \cite{chibowski2016zeta}, 
$x$ is the slip plane (\SI{0.24}{\nano\meter} \cite{chibowski2016zeta}), %\MO{Is it right definition? Maybe ``split plane thickness''?} \YP{Here is the definition from our reference papers. -> The slip plane is the plane defined by the distance at which the structure with its chemically bound water and ions moves in bulk through the solution as indicated in Eq. (19.2). It is the plane at which the zeta potential is valid. The zeta potential is the “modified” or “effective” surface charge. \href{https://www.sciencedirect.com/topics/engineering/slip-plane}{link}}, 
and $\zeta$ is the zeta potential. The measured average zeta potentials for the GUVs {and SUVs are} reported in Table \ref{tab:zeta}. 
The negative value for the GUVs is in line with other studies \cite{chibowski2016,MAKINO1991175}.
We hypothesize that the reason for it is the dipole rearrangement. %\anna{Sheereen, let me know
%of you want to elaborate further.}

\begin{table}[htb]
	\centering
	\begin{tabular}{|c|c|}
		\hline
		Vesicle & Zeta Potential \\
		\hline
		GUV &  -\SI{8.56}{\milli\volt} \\
		\hline
		PAT1 &  \SI{18.35}{\milli\volt} \\
		\hline
		PAT2 & \SI{18.87}{\milli\volt}  \\
		\hline
		PAT3  &  \SI{20.41}{\milli\volt} \\
		\hline
	\end{tabular}
	\caption{Measured average zeta potentials for the GUVs and SUVs used in the experiments.}\label{tab:zeta}
\end{table}

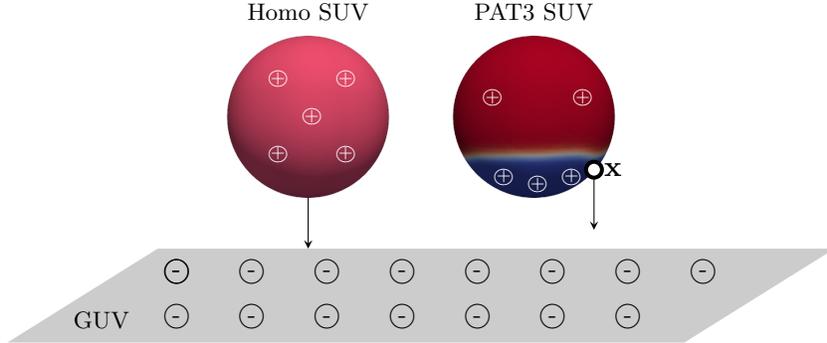
\begin{figure}[htb!]
	\centering
	\begin{tikzpicture}
		\fill[opacity=0.4, color=gray] (2,-2.5) -- (11,-2.5) -- (13,-1.25) -- (4,-1.25) --cycle;
		\coordinate[label=right:\textcolor{black}{ $\vect x$}] (P) at (9.8,-0.2);

		\draw [-stealth](9.8,-0.2) -- (9.8,-1);
		\draw [-stealth](6,-0.5) -- (6,-1.25);
		\node at (9,0.5) {\includegraphics[width=.2\linewidth]{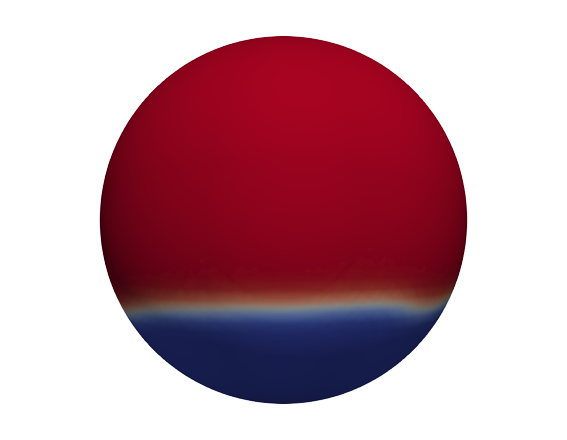}} ;
		\node at (6,0.5) {\includegraphics[width=.2\linewidth]{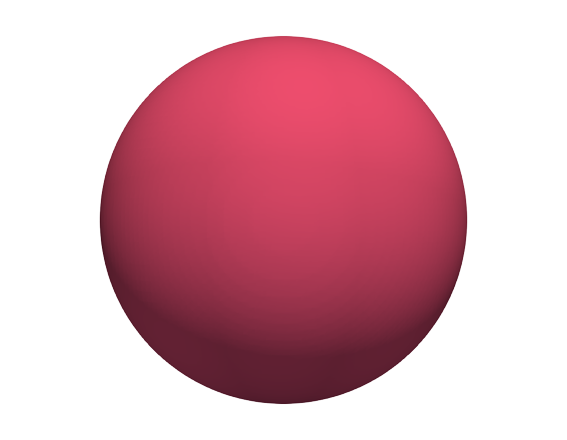}} ;
		%%%%%%%%%%%%%%%%%%%%%%%%%%%%%%%%%%%%%%%%%%%%%%%%%%%%%%%%%%%%%%%
		%Negative charges
		%!!!Remove if not needed
		\node at (3.25,-2.2) {\footnotesize  \textcolor{black}{GUV}} ;
		\node at (4.25,-2.2) {\footnotesize  \textcolor{black}{\textcircled{-}}} ;
		\node at (5.25,-2.2) {\footnotesize  \textcolor{black}{\textcircled{-}}} ;
		\node at (6.25,-2.2) {\footnotesize  \textcolor{black}{\textcircled{-}}} ;
		\node at (7.25,-2.2) {\footnotesize  \textcolor{black}{\textcircled{-}}} ;
		\node at (8.25,-2.2) {\footnotesize  \textcolor{black}{\textcircled{-}}} ;
		\node at (9.25,-2.2) {\footnotesize  \textcolor{black}{\textcircled{-}}} ;
		\node at (10.25,-2.2) {\footnotesize  \textcolor{black}{\textcircled{-}}} ;
		\node at (4.25,-1.6) {\footnotesize  \textcolor{black}{\textcircled{-}}} ;
		\node at (4.25,-1.6) {\footnotesize  \textcolor{black}{\textcircled{-}}} ;
		\node at (5.25,-1.6) {\footnotesize  \textcolor{black}{\textcircled{-}}} ;
		\node at (6.25,-1.6) {\footnotesize  \textcolor{black}{\textcircled{-}}} ;
		\node at (7.25,-1.6) {\footnotesize  \textcolor{black}{\textcircled{-}}} ;
		\node at (8.25,-1.6) {\footnotesize  \textcolor{black}{\textcircled{-}}} ;
		\node at (9.25,-1.6) {\footnotesize  \textcolor{black}{\textcircled{-}}} ;
		\node at (10.25,-1.6) {\footnotesize  \textcolor{black}{\textcircled{-}}} ;
		\node at (11.25,-1.6) {\footnotesize  \textcolor{black}{\textcircled{-}}} ;
		%!Charges on blue phase below
		\node at (9.0,1.9) {\footnotesize  \textcolor{black}{PAT3 SUV}} ;
		\node at (8.6,-0.3) {{\tiny \textcolor{white}{\textcircled{+}}}} ;
		\node at (9.5,-0.3) {{\tiny \textcolor{white}{\textcircled{+}}}} ;
		\node at (9.05,-0.4) {{\tiny \textcolor{white}{\textcircled{+}}}} ;
		
		%!Charges on red phase , less dense
		\node at (8.45,0.75) {{\tiny \textcolor{white}{\textcircled{+}}}} ;
		%\node at (9.05,1.1) {{\tiny \textcolor{white}{\textcircled{+}}}} ;
		\node at (9.65,0.75) {{\tiny \textcolor{white}{\textcircled{+}}}} ;
		%!Charges on pink phase, homogeneous 
		\node at (6.0,1.9) {\footnotesize  \textcolor{black}{Homo SUV}} ;
		\node at (5.6,-0.0) { \tiny \textcolor{white}{\textcircled{+}}} ;
		\node at (5.6,1) { \tiny \textcolor{white}{\textcircled{+}}} ;
		\node at (6.5,-0.0) {\tiny  \textcolor{white}{\textcircled{+}}} ;
		\node at (6.5,1) {\tiny  \textcolor{white}{\textcircled{+}}} ;
		\node at (6.05,0.5) {\tiny  \textcolor{white}{\textcircled{+}}} ;
		%%%%%%%%%%%%%%%%%%%%%%%%%%%%%%%%%%%%%%%%%%%%%%%%%%%%%%%%%%%%%%%
		\draw[color=black, fill=white, line width=1.5] (P) circle (3pt);
	\end{tikzpicture}
	\caption{
	Relative positions of GUV, represented as a plane, and a positively charged SUV, homogeneous (sphere on the left)  or
	phase-separated PAT3 SUV (sphere on the right), in a simulation.  The $L_o$ phase in 
	the phase-separated SUV is colored in red, while the $L_d$ phase is blue. 
%\MO{According to the data in tables it is \emph{disordered} (i.e. blue) phase that occupies 70\% of the SUV, am I right?}
%\anna{No, the area fraction $a_D$ refers to the liquid order domainds, i.e., the rafts (not the lakes).}
%\anna{Sheereen, are you ok with the negative signs on the GUV? I have added an explanation for it when we introduce Table \ref{tab:zeta}, 
%but if you're not ok with it we can change the figure.}
	}
\label{fig:force}
\end{figure}

Once the electric field $\textbf{E}$ is computed, the electrostatic force $\textbf{F}_e$ in \eqref{grache-1m} is given by
$\textbf{F}_e(\vect x) = \textbf{E} q(\vect x)$, where $q$ is a point charge located at
$\vect x$ on an SUV (see Fig.~\ref{fig:force}). Since we cannot measure a point charge on an SUV, we resort to 
an approximation. We find the surface charge density \eqref{grahame} for an SUV using the measured zeta potentials
reported in Table \ref{tab:zeta} for each composition under consideration. With the SUV surface charge density, 
we get the total attraction force density and we distribute it proportionally to the SUV surface. 
To exemplify the calculation, we consider a PAT3 SUV, which has $a_D = 70.37\%$, i.e., about 70\% of the surface of the SUV
is covered by the $L_o$ phase (red in Fig.~\ref{fig:force}). For composition PAT3, the concentration of DOTAP 
in the $L_d$ phase (blue in Fig.~\ref{fig:force}) is 41.8\% (see Table \ref{tab:distrib}), corresponding to 67.15\% of the total DOTAP in the SUV.  %\MO{I don't see how one gets 57.97\%. My calculations give 91.8\% of total DOTAP in disorder phase through   91.8\%= 0.418*0.7/(0.418*0.7+0.0861*0.3)*100\%} \YP{0.7 is liquid ordered phase, so it should multiply the other way 67.15\%= 0.418*0.3/(0.418*0.3+0.0861*0.7)*100\% , right? But it is still not 57.97, probably it was a typo, because I don't this number in any of my calculation files.}
So, we uniformly distribute 67.15\% of the total charge density, and hence force, to the $L_d$ phase. 
%\anna{Yerbol, have I understood it right or am I missing something? Somehow, I don't see where you use this:
%We compute the total (not pointwise) electrostatic force between the GUV and a homogeneous SUV
%and assume that the force between the GUV and a phase separated SUV is the same.}\YP{Force formula gives the total force between vesicles, it doesn't take into account the phase separation. Since pointwise force is different depending on phase, we have to distribute it accordingly. According to our calculations, total attraction force is same whether it is homogeneous or phase separating. So fusion depends more on how force is distributed on the surface, rather than the total amount of force.}

Problem \eqref{grache-1m}--\eqref{gracke-4} needs to be
supplemented with initial values of velocity $\bu_0$ and state $c_0$. 
We take $\bu_0 = \boldsymbol{0}$ (surface fluid at rest) and $c_0$ corresponding 
to a homogeneous mixture. We define $c_0$ as a realization of Bernoulli random variable~$c_\text{rand} \sim \text{Bernoulli}({a_D})$
with mean value domain area fraction ${a_D}$.
%\begin{equation}\label{raftIC}
%	 \coloneqq c_\text{rand}(\vect x)\quad\text{for active mesh nodes $\vect x$}.
%\end{equation}
The value of ${a_D}$ is set according to the thermodynamic principles described in \cite{zhiliakov2021experimental}, 
which align the values to the measured quantities reported in Table \ref{tab:comp}. %\MO{Table~2?}.
%\anna{Should we say any more?}

In generic settings, the solutions to the NSCH problem can only be found numerically. 
Our numerical scheme for problem~\eqref{grache-1m}-\eqref{gracke-4} relies on 
an unfitted finite element method called Trace FEM~\cite{olshanskii2017trace2} and 
an adaptive time stepping technique~\cite{gomez2008isogeometric}. A thorough
description of our methodology can be found in \cite{WANG2022183898}, with
more details available in \cite{Palzhanov2021,yushutin2020numerical,Yushutin_IJNMBE2019}.
We performed a mesh refinement study to identify a mesh that yields
approximations of $\bu$, $p$, $c$, and $\mu$ (denoted with $\bu_h$, $p_h$, $c_h$, $\mu_h$) 
with a satisfying level of accuracy. For the results in Sec.~\ref{sec:res}, we 
adopted mesh with 225822 active degrees of freedom (193086 for $\bu_h$
and 10912 for $p_h$, $c_h$, and $\mu_h$). The time step $\Delta t$ adaptively varies
from  $\Delta t=$4$\cdot 10^{-6}$ s during the fast initial phase of spinodal decomposition to about $\Delta t=$8$\cdot 10^{-4}$ s
during the later slow phase of lipid domain coarsening and growth, and up to $\Delta t= 4$ s when the process is close to equilibrium.
%\anna{Yerbol, please update if this is not right.} \YP{Right.}

We recall that our numerical method produces numerical solutions that satisfy the mass conservation 
principle behind \eqref{grache-1m}-\eqref{gracke-4}:
\begin{equation}\label{raftFracDiscrete}
\int_{\Gamma} c_h(\vect x, t_n) \diff{s}=\int_{\Gamma} c_h(\vect x, t_{n-1}) \diff{s}\quad\text{implying}\quad	\frac{\int_{\Gamma} c_h(\vect x, t_n) \diff{s}}{\int_{\Gamma} 1 \diff{s}} \simeq {a_D},
\end{equation}
for all $n=1,\dots,N$. 

\section{Results and Discussion}\label{sec:res}

In order to investigate the effect of surface density of cationic lipid DOTAP on liposomes' fusogenicity, 
we selected a phase-separating lipid composition DOPC:DPPC:Chol and focused on three different molar ratios 
reported in Table \ref{tab:comp} with distinct domain ($L_o$) area fractions $a_D$ listed in Table \ref{tab:physical_param}. 
We replaced 15 mol\% of DOPC in these liposomal formulations with DOTAP. Given that DOTAP's acyl-chain 
chemistry is similar to that of DOPC, we assumed that this lipid would have similar phase partitioning behavior as 
DOPC and would mostly partition into the $L_d$ phase. 
Table \ref{tab:distrib} summarizes the lipid distribution among $L_o$ and $L_d$ phases. 
These lipid distributions are estimated based on the tie-lines available in literature \cite{Veatch17650} 
and as described in our previous studies \cite{WANG2022183898,zhiliakov2021experimental}. With the same 
DOTAP content, composition PAT3 is expected to have the highest surface density of DOTAP in $L_d$ phase
because it has the largest $a_D$, 
and composition PAT1 is expected to have the lowest density of DOTAP in its $L_d$ phase because it has the smallest $a_D$. 

\begin{table}[htb]
\centering
\begin{tabular}{ |c|c|c|c|c|c|c|c|c|}
 \hline 
 & \multicolumn{4}{| c |}{\footnotesize{$L_d$ phase}}  & \multicolumn{4}{| c |}{
 \footnotesize{$L_o$ phase}} \\
  \hline 
 \footnotesize{Composition}&  \footnotesize{DOTAP}  &  \footnotesize{DOPC} & \footnotesize{DPPC} & \footnotesize{Chol} & \footnotesize{DOTAP}  &  \footnotesize{DOPC} & \footnotesize{DPPC} & \footnotesize{Chol}  \\
  \hline
 PAT1 (15\%) & 16.67\% & 49.33\% & 16\% & 18\% & 5.56\% & 16.44\% & 43\% & 35\% \\
 \hline
 PAT2 (15\%) & 22.91\% & 41.09\% & 29\% & 7\% & 4.65\% & 8.35\% & 61\% & 26\% \\
\hline
 PAT3  (15\%) & 41.80\% & 26.20\% & 24\% & 8\% & 8.61\% & 5.39\% & 57\% & 29\% \\
 \hline
\end{tabular}
\caption{Lipid distribution among the two phases in the examined phase-separated SUVs.}\label{tab:distrib}
\end{table}%

\begin{figure}[htb]
	\centering
\begin{overpic}[width=.6\textwidth,grid=false]{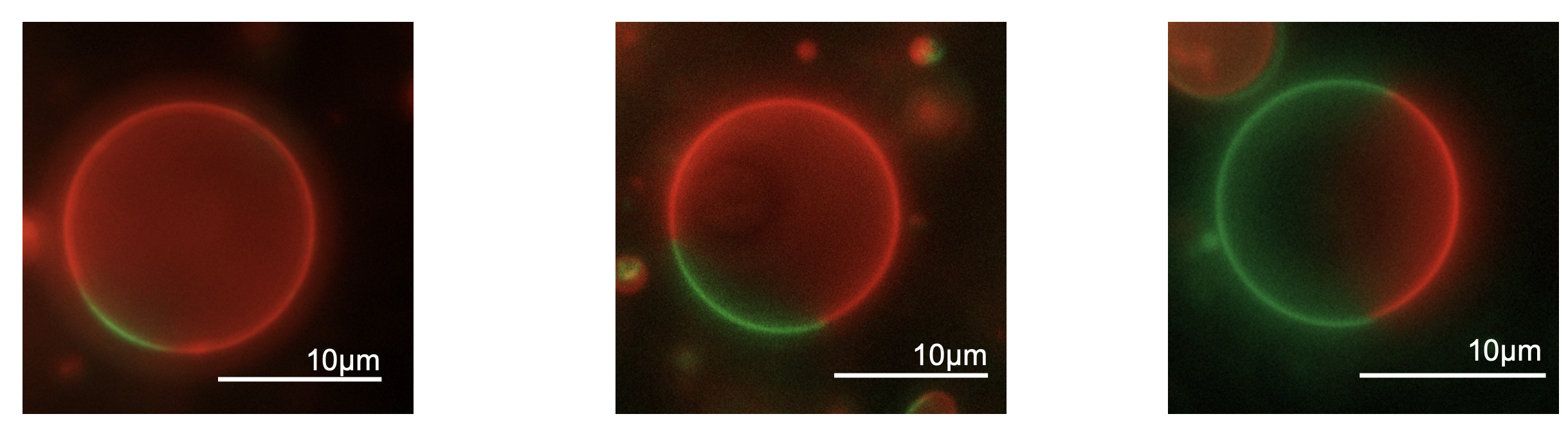}
			\put(5,27){\small{PAT1 (15\%)}}
			\put(42,27){\small{PAT2 (15\%)}}
			\put(78,27){\small{PAT3 (15\%)}}
		\end{overpic}
	\caption{Fluorescence microscopy images of representative 
	phase-separated GUVs with composition PAT1 (left), PAT2 (center), and PAT3 (right).
	Red fluorescence shows $L_d$ phase and green fluorescence shows $L_o$ phase.}
	\label{fig:steady_state}
\end{figure}

To confirm that the partial replacement of DOPC with DOTAP does not interfere 
with phase separation in the examined lipid compositions, we first prepared GUVs of these formulations 
because these micron-sized liposomes can be visualized under optical microscopy.  Fig.~\ref{fig:steady_state} 
depicts epifluorescent images of representative GUVs with lipid compositions tested here, 
where the red patches are $L_d$ phase and the green patches are $L_o$ phase. 
These images confirmed that the membrane phase separation occurred as expected in all three examined compositions.
The results were in good agreement with our previous findings reported in \cite{WangMajd2023}.

\vskip .3cm

Next, we prepared SUVs with the above-mentioned compositions to study their fusogenicity. 
These phase-separating liposomes were compared to homogenous liposomes composed of DOPC 
with different amounts of DOTAP. We evaluated these SUVs for size and zeta potential. 
Dynamic light scattering %(DLS) 
measurements showed that the size distribution of SUVs 
had a reduction when DOTAP was included in the formulation and was comparable among the 
three DOTAP-containing phase-separating compositions (PAT1, PAT2, and PAT3) (Fig.~SI 1A). 
The zeta potential values in homogeneous SUVs increased with an increase in their DOTAP content and were similar
in all three phase-separating compositions. Interestingly, phase-separating SUVs showed slightly
higher zeta potential compared to homogenous SUVs with same DOTAP content (Fig.~SI 1B), 
presumably due to the asymmetrical charge distribution on these SUVs that has been reported 
to affect the zeta potential values \cite{Pardhy2010,WangMajd2023}. 

\begin{figure}[htb!]
	\centering
	\includegraphics[width = .6\textwidth]{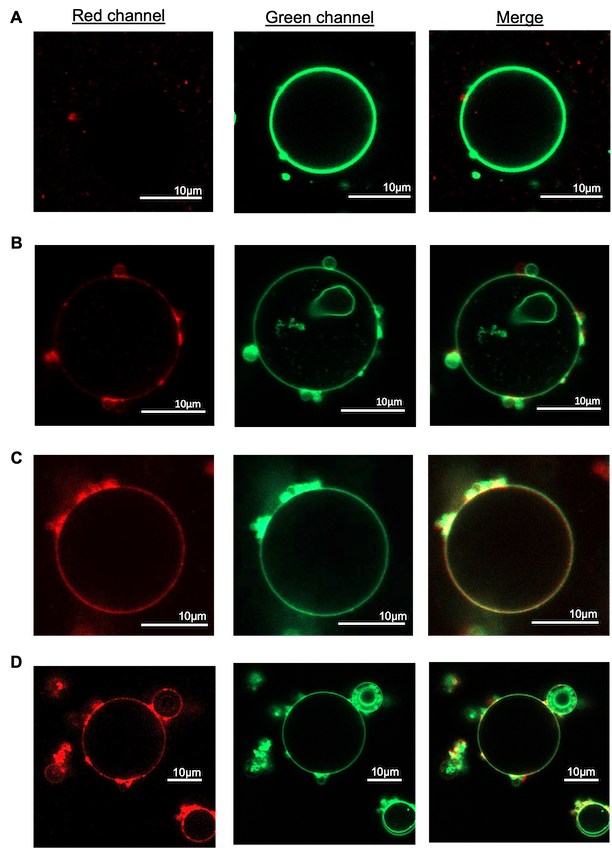}
	\caption{Fluorescence confocal images of representative GUVs composed of DOPC labeled with 0.3\% green fluorescent AF488-PE after 10 min incubation with SUVs of (A) Homo (0\% DOTAP), (B) Homo (15\% DOTAP), (C) Homo (30\% DOTAP), (D) PAT3 (15\% DOTAP).}
	\label{fig:fluore}
\end{figure}

To examine the ability of DOTAP-containing phase-separating SUVs to fuse into other membranes, 
we incubated them with GUVs of DOPC composition at 37\textdegree{}C for 10 min. 
After the incubation, samples were imaged with confocal microscopy to evaluate the level of fusion of SUVs 
(labeled with red fluorescence) into GUVs (labeled with green fluorescence). In case of homogenous SUVs with no DOTAP, 
the GUVs exhibited only green fluorescence indicating no significant fusion (Fig.~\ref{fig:fluore}A). 
Increasing DOTAP concentration to 15\% in homogenous SUVs resulted in a mixture of both 
red and green fluorescence on GUVs suggesting some level of fusion (Fig.~\ref{fig:fluore}B). 
Further increasing DOTAP to 30\% led to a stronger red fluorescence signal, indicating higher level of fusion (Fig.~\ref{fig:fluore}C). 
Interestingly, incubation of GUVs with phase-separating SUVs of PAT3 composition (with 15\% DOTAP), 
led to much stronger red fluorescence signal in GUV membranes compared to that in case of homogenous liposomes with 
15\% DOTAP, and was comparable to that of homogenous SUVs with 30\% DOTAP. 

\begin{figure}[htb!]
	\centering
	\includegraphics[width = .48\textwidth]{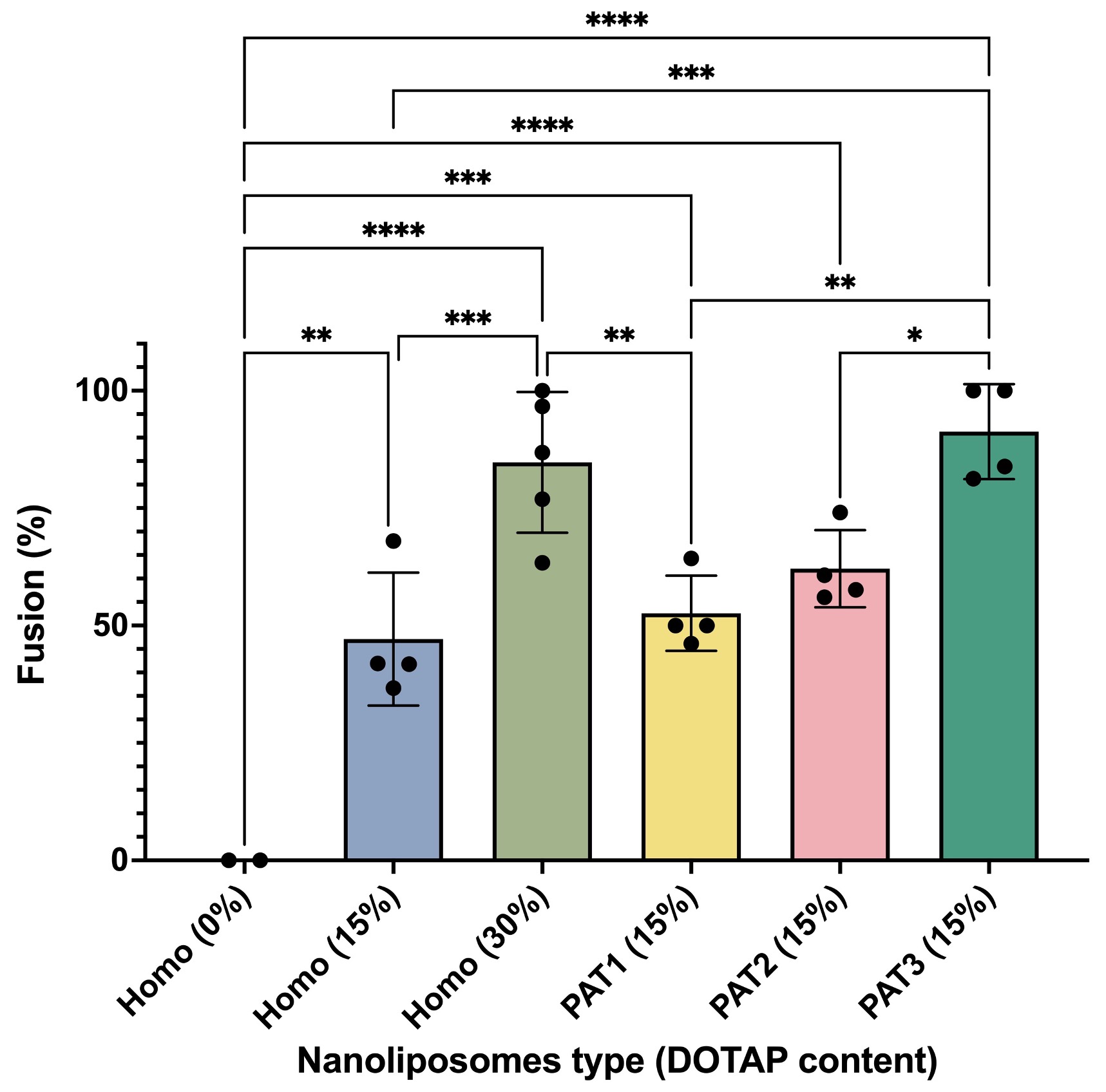}
	\caption{Fraction of GUVs that showed fusion after \SI{10}{\minute} incubation with SUVs of different lipid compositions.
Data points represent the fraction of GUVs from independent experiments. Error bars correspond to the standard deviation. 
Data were statistically analyzed using one-way ANOVA and $^*$: p value$<$0.05, $^{**}$: p value$<$0.01, $^{***}$: p values$<$0.001, $^{****}$: p value$<$0.0001.
}
	\label{fig:fusion_frac}
\end{figure}

To quantify the level of fusion in these experiments, we measured the fraction of 
GUVs that showed fusion upon incubation with SUVs. As summarized in Fig.~\ref{fig:fusion_frac}, 
higher DOTAP concentration resulted in higher level of fusion and PAT3 composition, 
with highest DOTAP density in $L_d$ phase, showed the highest level of fusion. These results showed 
that increasing the surface density of DOTAP on SUVs even locally (through phase separation) can enhance their fusogenicity.

\begin{figure}[htb!]
	\centering
%	\begin{tikzpicture}  
%		\begin{axis}  
%			[  
%			ybar, 
%			bar width= 40,
%			enlargelimits=0.15,
%			ymajorgrids = true,  
%			ylabel={Minutes}, % the ylabel must precede a # symbol.  
%			xlabel={\ lipid domain area fraction},  
%			symbolic x coords={10.8\%, 34.47\%, 70.37\% }, % these are the specification of coordinates on the x-axis.  
%			xtick=data,  
%			nodes near coords, % this command is used to mention the y-axis points on the top of the particular bar.  
%			nodes near coords align={above,color = blue},  
%			]  
%			\addplot[red!20!black,fill=red!80!white] coordinates {(10.8\%,10) (34.47\%,19) (70.37\%,23.3)  };  
%		\end{axis}  
%	\end{tikzpicture} 
\begin{overpic}[width=.48\textwidth,grid=false]{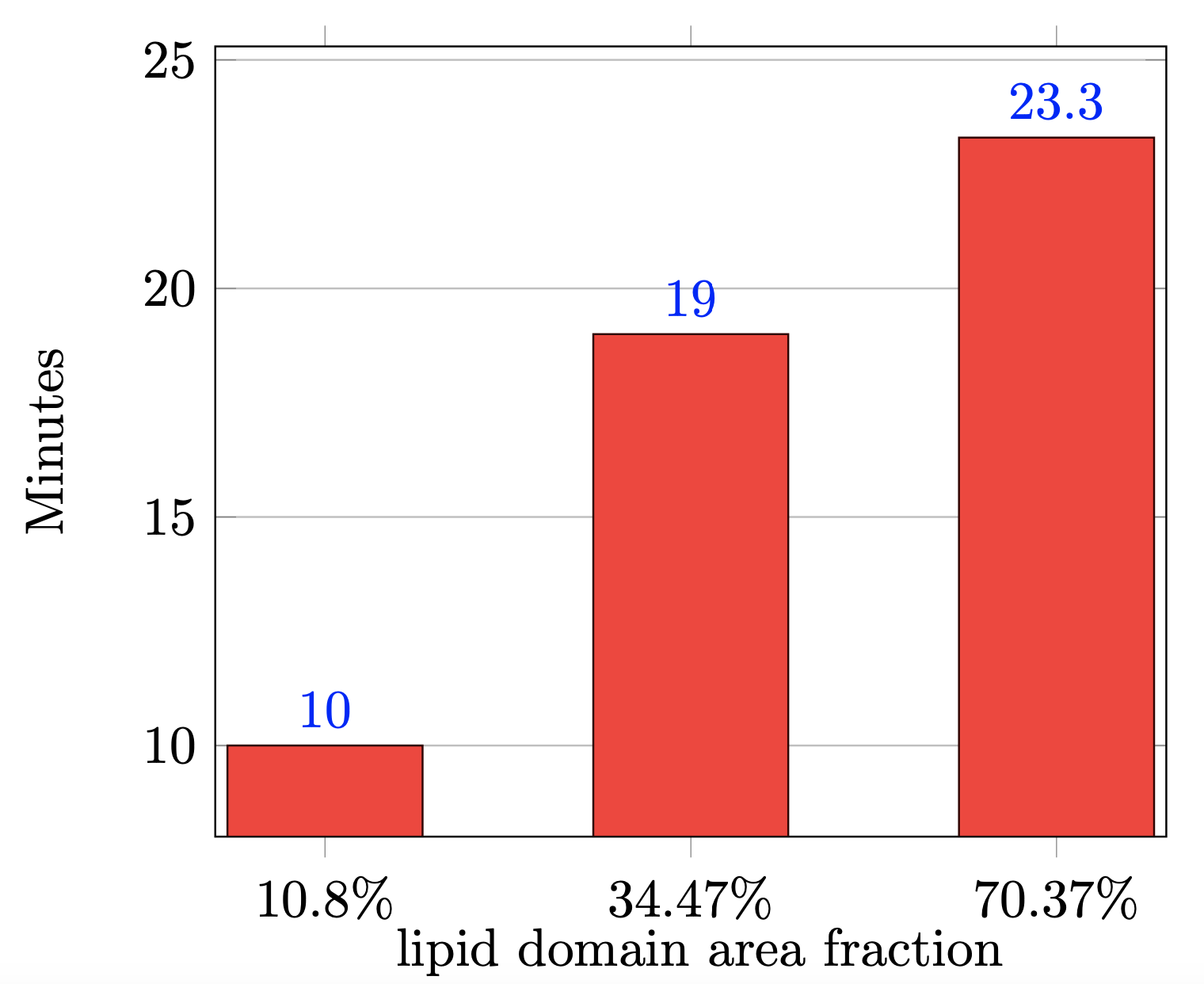}
			\put(22,15){\textcolor{blue}{PAT1}}
			\put(52,32){\textcolor{blue}{PAT2}}
			\put(82,40){\textcolor{blue}{PAT3}}
		\end{overpic}
	\caption{Average time needed for a simulated SUV to reach the equilibrium state (i.e., one patch of the minority phase
against the background of the majority phase) for the three compositions under consideration.}
	\label{fig:ave_steady_state}
\end{figure}

Next, we present the computational data and show how they corroborate the observations made 
from the experiments.
As mentioned in Sec.~\ref{sec:comp}, in phase-separated SUVs with cationic lipids there is a complex 
interplay of the forces driving phase separation, forces driving surface flow, and electrostatic forces.
In order to facilitate our understanding of how patches of fusogenic lipids promote fusion,
we let the SUVs undergo phase-separation before exposing them to the target model membranes
both in the simulations and in the experiments. This serves the purpose of
disentangling the effect of phase separation forces from the effect of electrostatic forces.
By the time the SUVs are exposed to the model membranes (i.e., $>60$ \SI{}{\minute} after formation), 
most SUVs have reached the equilibrium phase-separated state, mostly with one patch of the minority phase
against the background of the majority phase. %\anna{Check that the statement is correct.}
From our previous work \cite{WANG2022183898}, we know that membranes of different lipid compositions take different times
to reach the equilibrium state, specifically it happens faster for compositions with 
smaller $L_o$ domain area fractions.
See Fig.~\ref{fig:ave_steady_state} for the average time needed to reach the equilibrium 
for the three lipid compositions under consideration. The average is taken over five simulations
with the given composition and random initial distributions (as explained in Sec.~\ref{sec:comp}). 
We see that a PAT3 SUV ($a_D \approx 70\%$) takes more than the double of the time a PAT1 SUV ($a_D \approx 11\%$) needs
to reach the equilibrium state. 

%\begin{figure}%[htb!]
%	\centering
%	\includegraphics[width = .48\textwidth]{img/number_of_patches_all.pdf}
%	\begin{tikzpicture}  
%		\begin{axis}  
%			[  
%			ybar, 
%			bar width= 40,
%			enlargelimits=0.15,
%			ymajorgrids = true,  
%			ylabel={Minutes}, % the ylabel must precede a # symbol.  
%			xlabel={\ Composition area fraction},  
%			symbolic x coords={10.8\%, 34.47\%, 70.37\% }, % these are the specification of coordinates on the x-axis.  
%			xtick=data,  
%			nodes near coords, % this command is used to mention the y-axis points on the top of the particular bar.  
%			nodes near coords align={above,color = blue},  
%			]  
%			\addplot[red!20!black,fill=red!80!white] coordinates {(10.8\%,10) (34.47\%,19) (70.37\%,23.3)  };  
%		\end{axis}  
%	\end{tikzpicture} 
%	\caption{Number of patches \YP{I didn't find this illustration good enough for two reasons: 1-) the separation between compositions is not clearly seen, because during fusions of patches there is kind of race between 70\% and 34\% compositions, but 34\% reaches the steady state faster on average. 2-) in the current set up for 10.8\% it is fast enough that when patches start forming we can observe maximum 2 patches, sometimes 3 patches. So there are not many data points to add on the top of this graphs.}}
%	\label{number_of_rafts}
%\end{figure}

\begin{figure}[htb!]
	\begin{center}
		\href{https://www.youtube.com/watch?v=sFW1SaPRUzk}{\begin{overpic}[width=.13\textwidth,grid=false]{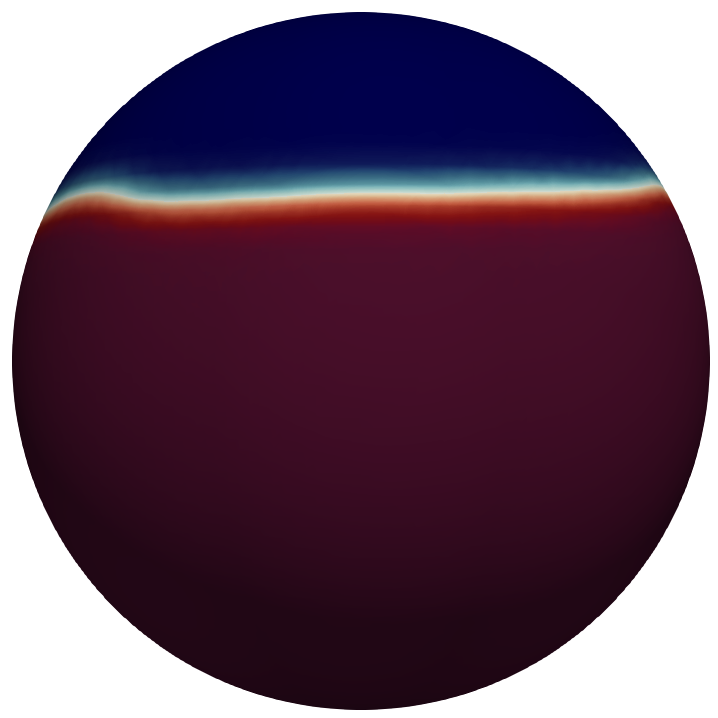}
			\put(30,102){\small{$t = 0$}}
			\put(-70,50){\small{PAT3}}
			\put(-93,32){\small{$a_D = 70.37$\%}}
		\end{overpic}
		\begin{overpic}[width=.13\textwidth,grid=false]{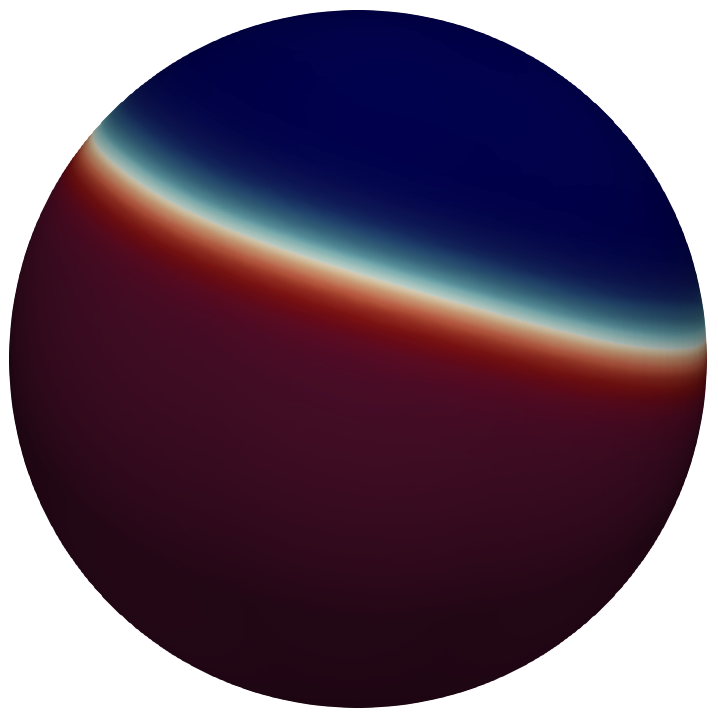}
			\put(30,100){\small{$t = 3$}}
		\end{overpic}
		\begin{overpic}[width=.13\textwidth,grid=false]{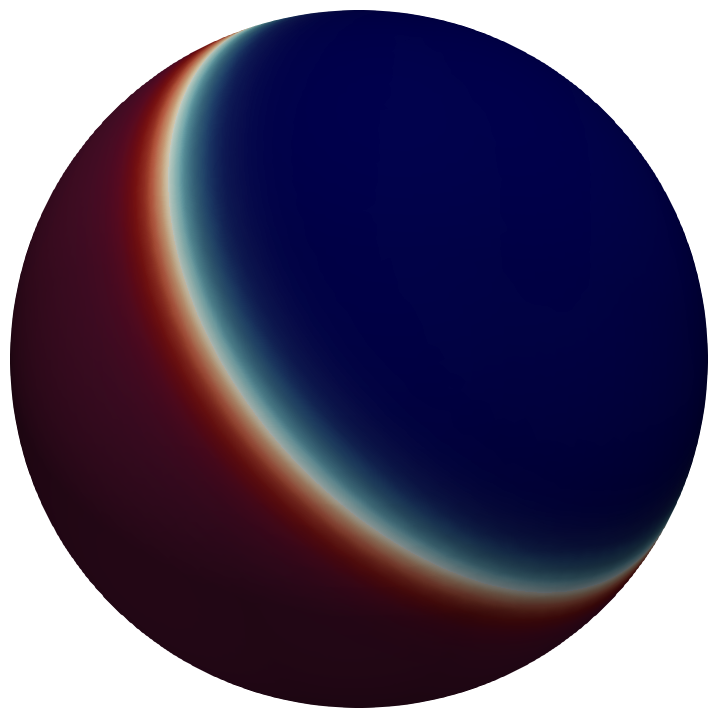}
			\put(30,102){\small{$t = 4$}}
		\end{overpic}
		\begin{overpic}[width=.13\textwidth,grid=false]{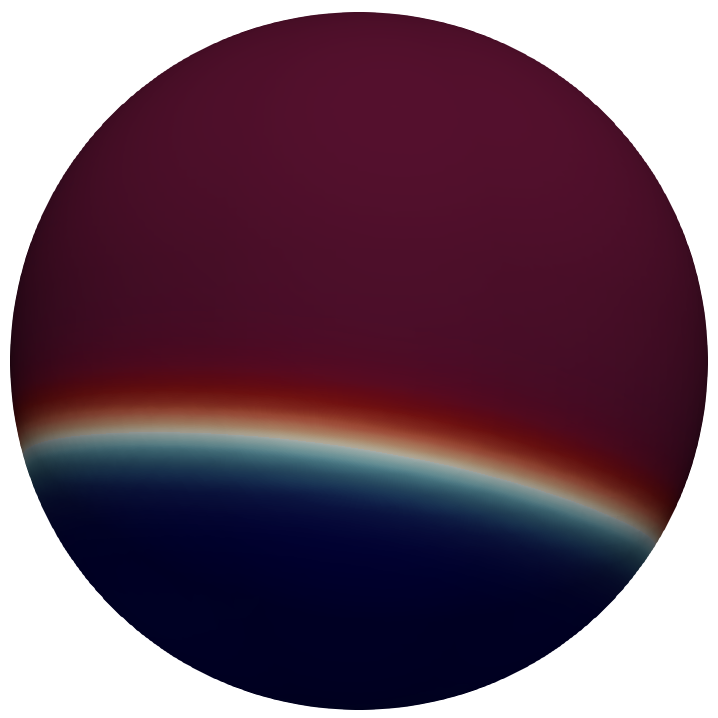}
			\put(30,102){\small{$t = 6$}}
		\end{overpic}}\\
		
		\vskip 12pt
		\href{https://www.youtube.com/watch?v=xy1OKYqablg}{\begin{overpic}[width=.13\textwidth,grid=false]{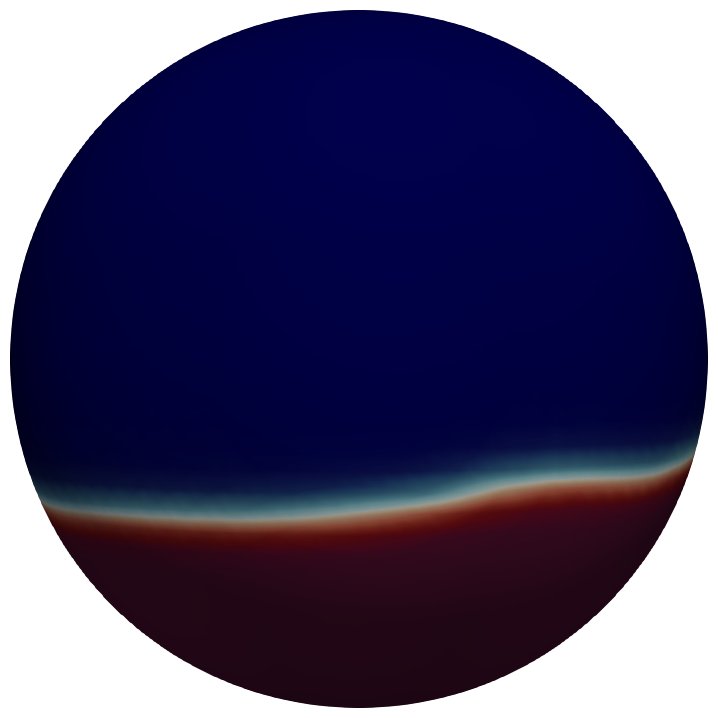}
			\put(30,102){\small{$t = 0$}}
			\put(-70,50){\small{PAT2}}
			\put(-93,32){\small{$a_D = 34.47$\%}}
		\end{overpic}
		\begin{overpic}[width=.13\textwidth,grid=false]{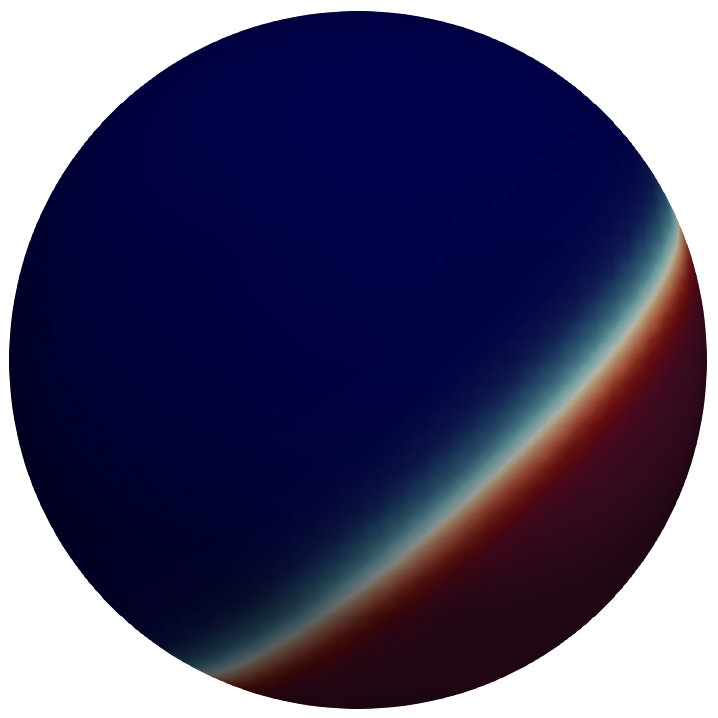}
			\put(30,100){\small{$t = 6$}}
		\end{overpic}
		\begin{overpic}[width=.13\textwidth,grid=false]{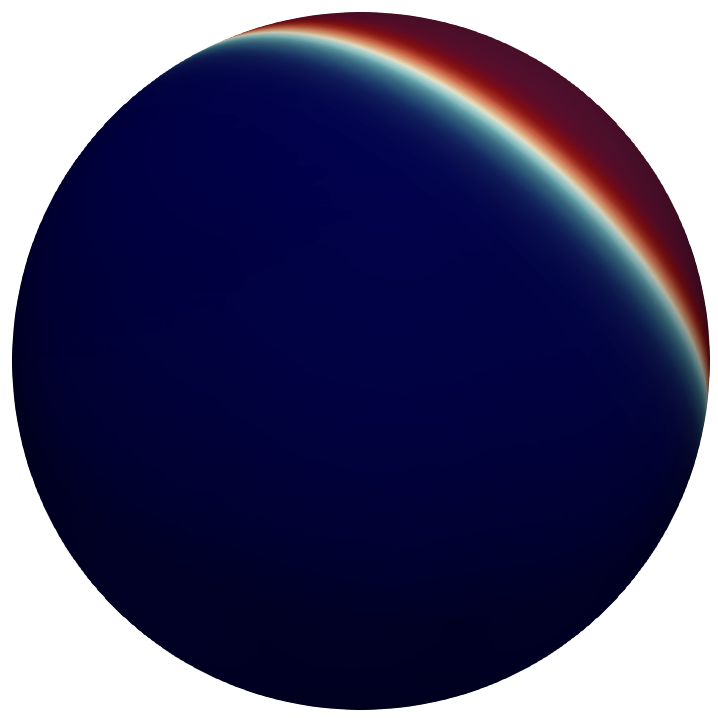}
			\put(30,102){\small{$t = 10$}}
		\end{overpic}
		\begin{overpic}[width=.13\textwidth,grid=false]{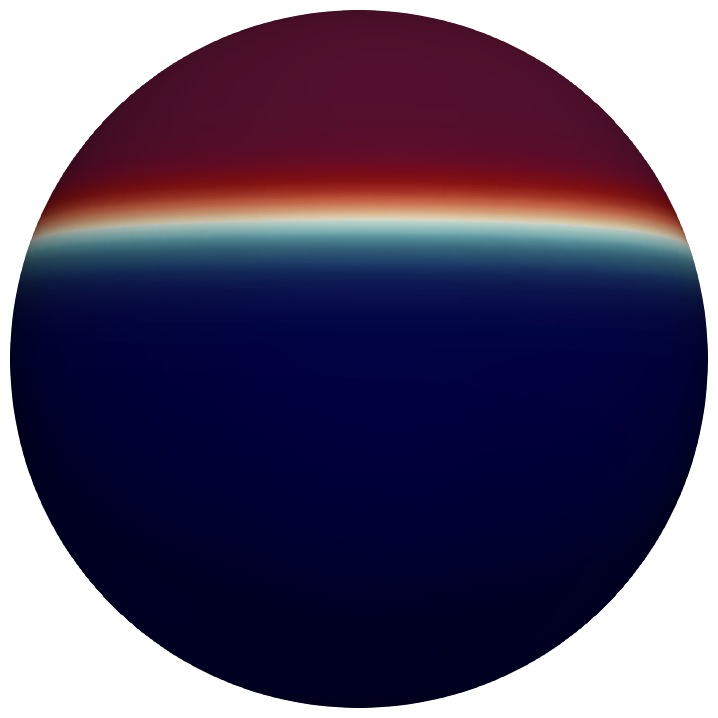}
			\put(30,102){\small{$t = 15$}}
		\end{overpic}}\\
		
		\vskip 12pt
		\href{https://www.youtube.com/watch?v=xy1OKYqablg}{\begin{overpic}[width=.13\textwidth,grid=false]{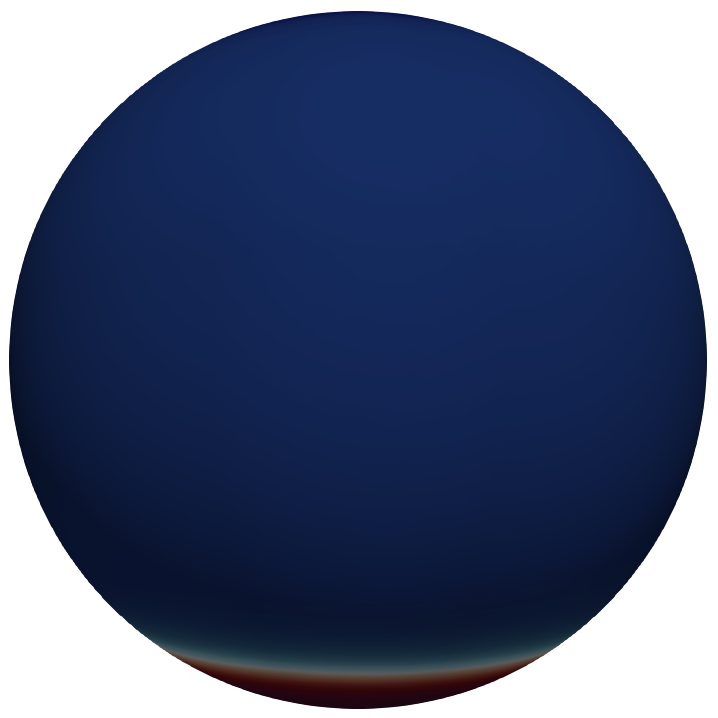}
			\put(30,102){\small{$t = 0$}}
			\put(-70,50){\small{PAT1}}
			\put(-93,32){\small{$a_D = 10.8$\%}}
		\end{overpic}
		\begin{overpic}[width=.13\textwidth,grid=false]{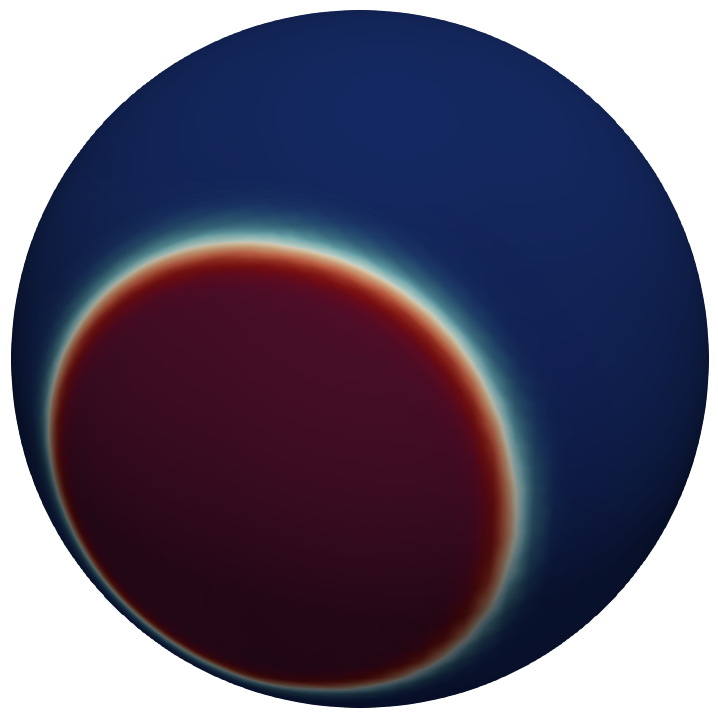}
			\put(30,100){\small{$t = 20$}}
		\end{overpic}
		\begin{overpic}[width=.13\textwidth,grid=false]{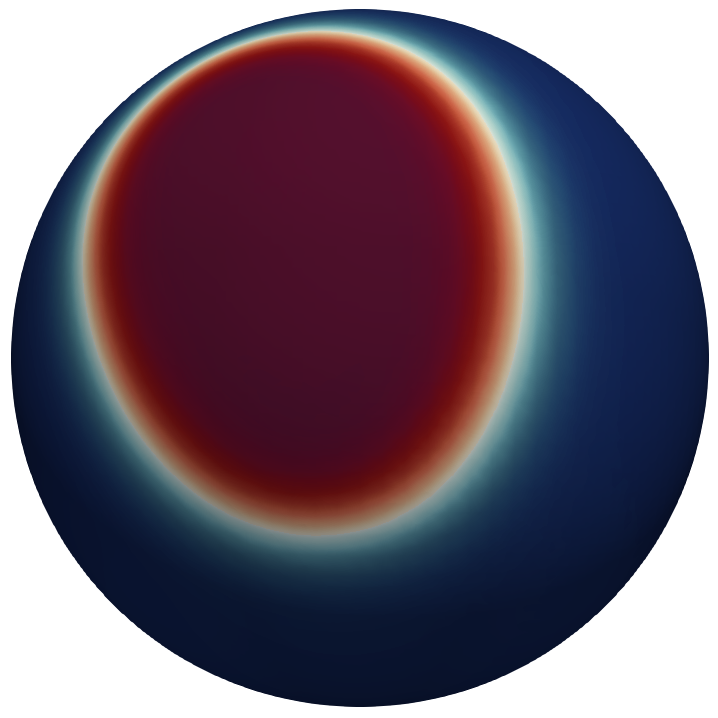}
			\put(30,102){\small{$t = 40$}}
		\end{overpic}
		\begin{overpic}[width=.13\textwidth,grid=false]{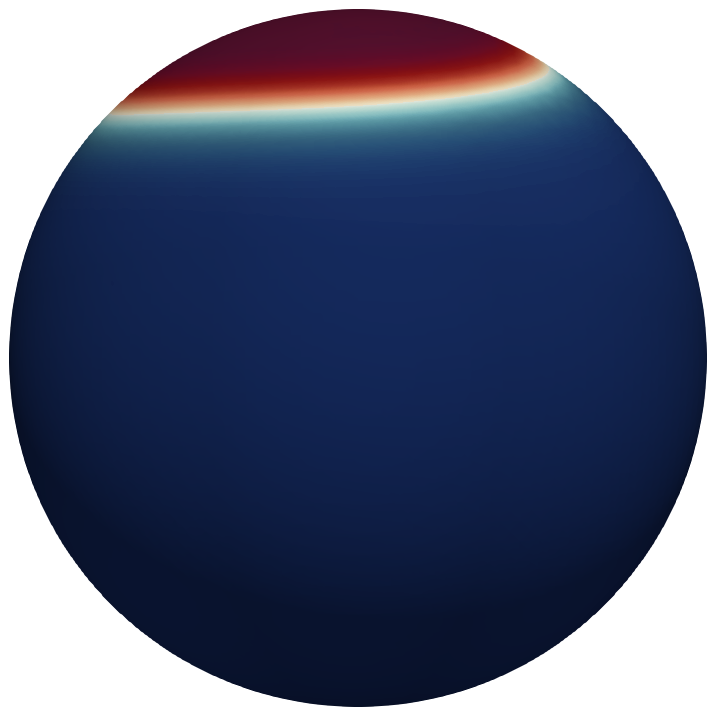}
			\put(30,102){\small{$t = 60$}}
		\end{overpic}}
		\begin{overpic}[width=.5\textwidth,grid=false]{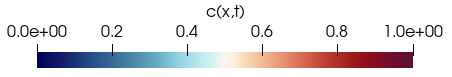}
		\end{overpic}

	\end{center}
	\caption{Snapshots of a simulation with the phase-separated PAT3 SUV (top), PAT2 SUV (center), and PAT1 SUV (bottom)
	at different times (min). 
	Red corresponds to the $L_o$ phase and blue to the $L_d$ phase. For each composition, the 
	$L_d$ phase is initially placed at the top of the SUV (first column).  The model membrane, not seen in the figure, 
        is represented as a horizontal plane below the SUV.
	Click any picture above to run the corresponding full animation. }
	\label{fig:numerical_snapshots}
\end{figure}

\begin{figure}[htb!]
\centering
\begin{overpic}[width=.48\textwidth,grid=false]{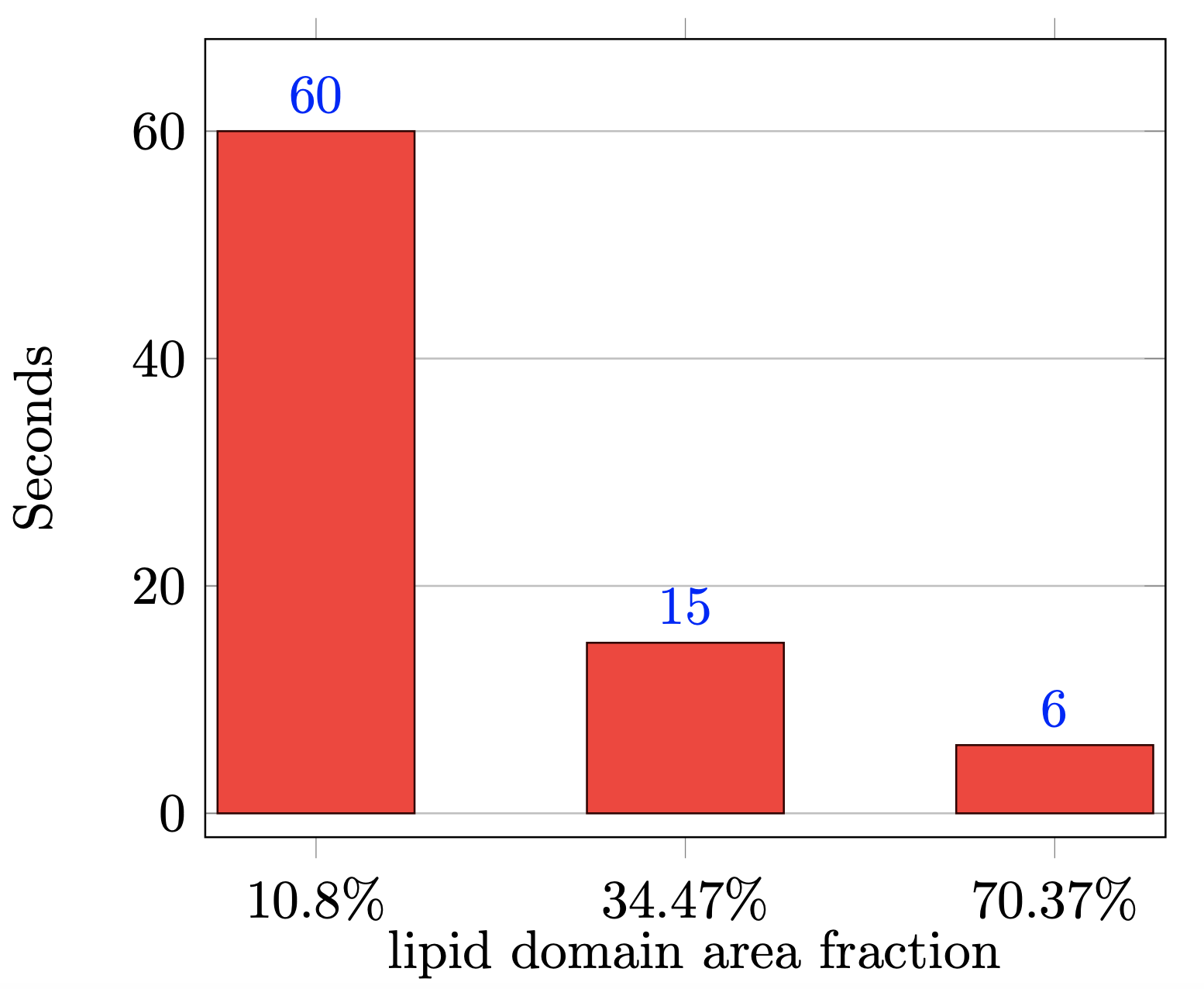}
			\put(21,42){\textcolor{blue}{PAT1}}
			\put(51.5,20){\textcolor{blue}{PAT2}}
			\put(82.5,15){\textcolor{blue}{PAT3}}
		\end{overpic}
	\caption{Average time needed to have the $L_d$ phase in a simulated SUV face the target model membrane
	starting from the worst-case scenario configuration. }
	\label{fig:time_to_reorient}
\end{figure}

Once a SUV has reached the phase-separated equilibrium, it is exposed to the target model membrane 
(equivalent of GUV in experiments), which is represented as a horizontal plane below the SUV in the simulations. 
Initially, we place the $L_d$ phase, which is the
phase with the majority of the positive charge, opposite to the model membrane, i.e., at the top of the SUV. 
See the first column in Fig.~\ref{fig:numerical_snapshots}. In a sense, this is the worst-case scenario as it will take 
the longest to reorient the $L_d$ phase so that it faces the model membrane. Once the $L_d$ phase
faces the model membrane, the SUV is in the optimal configuration to initiate fusion since the majority of
the fusogenic lipids is in the $L_d$ phase (see Table \ref{tab:distrib}). Fig.~\ref{fig:numerical_snapshots} shows
snapshots of the simulated reorientation process for the three compositions. 
From Fig.~\ref{fig:numerical_snapshots}, we clearly see that each SUV takes a different amount of time to have the 
$L_d$ phase face the model membrane. Fig.~\ref{fig:time_to_reorient} reports such (average) time
for each composition. The average is computed again over 5 simulations per composition, as explained above. 
We take this time as a proxy for the promotion of fusion since it is the time need to have 
the SUV in the optimal configuration for fusion, i.e., with the majority of the fusogenic lipids facing the GUV. 
%\anna{Is this clear or should I add more?} \MO{Sounds clear for me. -M.}
Fig.~\ref{fig:time_to_reorient} informs us that in average a PAT1 SUV takes ten times longer 
than a PAT3 SUV to reorient its $L_d$ phase. Recall that the data used for Fig.~\ref{fig:fusion_frac} were acquired after
10 min of incubation. In that amount of time, the simulations predict that all PAT3 SUVs were in the optimal configuration 
for fusion, regardless of the initial position of the $L_d$ phase with respect to the GUV. In contrast, the PAT1 and PAT2 SUVs exposed to a GUV
in the worst-case scenario (i.e., $L_d$ phase opposite to the GUV) did not have sufficient time to have the $L_d$ phase face the GUV. 
This provides an explanation why the PAT3 SUVs outperform both the PAT1 and PAT2 SUVs. 

%\begin{figure}[htb!]
%	\centering
%	\begin{tikzpicture}  
%		\begin{axis}  
%			[  
%			ybar, 
%			bar width= 40,
%			enlargelimits=0.15,
%			ymajorgrids = true,  
%			ylabel={Seconds}, % the ylabel must precede a # symbol.  
%			xlabel={\ lipid domain area fraction},  
%			symbolic x coords={10.8\%, 34.47\%, 70.37\% }, % these are the specification of coordinates on the x-axis.  
%			xtick=data,  
%			nodes near coords, % this command is used to mention the y-axis points on the top of the particular bar.  
%			nodes near coords align={above,color = blue},  
%			]  
%			\addplot[red!20!black,fill=red!80!white] coordinates {(10.8\%,60) (34.47\%,15) (70.37\%,6)  };  
%		\end{axis}  
%	\end{tikzpicture}  
%	\caption{Time to reorient (Worst-case scenario)}
%	\label{time_chart}
%\end{figure}

\section{Conclusions}
Fluorescence microscopy studies of DOTAP-enriched SUVs fusing with GUVs having a DOPC membrane reveal two main observations:
(i) The addition of DOTAP enhances fusogenicity of SUVs, and the fusogenicity levels increase 
with higher percentages of DOTAP in the SUV composition,
(ii) DOTAP-charged SUVs with a phase-separated membrane exhibit higher fusogenicity levels compared to DOTAP-charged SUVs with a homogeneous membrane. Notably, stronger fusogenicity was observed for SUVs with higher concentrations of DOTAP in smaller patches of the membrane in the liquid disordered phase.

While the first observation was expected due to the positive charge carried by DOTAP and the (slightly) negative potential measured for the DOPC GUVs, the second observation is more intriguing. We propose the following explanation for (ii): the phase separation leads to a higher local positive charge density on the SUV membrane, enhancing its interactions with the target GUV membrane. Moreover, the formation of patches and favorable orientation occurs more rapidly for lipid compositions with area fraction of the liquid disordered phase. Computational results using a state-of-the-art continuum-based model of the two-phased fluid membrane support this suggestion.
Findings of this study can be applied for the design of highly fusogenic delivery cationic liposomes with minimal levels of toxicities.
%\MO{Edit and update.}

\section*{Funding information}
\noindent  This work was partially supported by US National Science Foundation (NSF) through grant DMS-1953535.
M.O.~acknowledges the support from NSF through DMS-2309197.
S.M.~acknowledges the support from NSF through DMR-1753328.
% Uncomment if using bibtex (default)

\section*{Author Contributions} 

\noindent  {\bf Y.~Wang}:  design and conducting experiments, data collection and analysis, writing -- original draft, review; {\bf Y.~Palzhanov}: software, analysis of the computational data, visualization; 
{\bf D.~T.~Dang}:  conducting experiments, data collection; {\bf A.~Quaini}: supervision of computational studies, 
writing -- original draft, review and editing;
{\bf M.~Olshanskii}: methodology, mathematical modeling,  writing -- review and editing; {\bf S.~Majd}: supervision of all experimental studies, writing -- original draft, review, and editing.

\section*{Conflict of Interest and other Ethics Statements}

\noindent The authors report no conflict of interest. 

\bibliographystyle{plain}
\bibliography{literatur}

\pagebreak

\begin{center}
\Large{Supplementary Information}
\end{center}

\noindent {\bf Size distribution and zeta potential of homogenous and phase-separating SUVs}

Homogenous SUVs composed of DOPC with different amounts of DOTAP and phase-separating SUVs (compositions PAT1, PAT2 and PAT3) with 15 mol\% DOTAP were prepared and evaluated for size distribution and zeta potential using Malvern Zetasizer. 

As shown in Figure S1A, dynamic light scattering data revealed that the average size of all examined SUV formulations was within 120-150 \SI{}{\nano\meter} range. The zeta potential of homogenous SUVs increased with an increase in their DOTAP content, as expected.  With the same DOTAP content, phase-separating SUVs had slightly higher zeta potential compared to homogenous SUVs, as summarized in Figure S1B.

\begin{figure}[htb!]
\renewcommand\thefigure{S1}
	\centering
	\includegraphics[width = .8\textwidth]{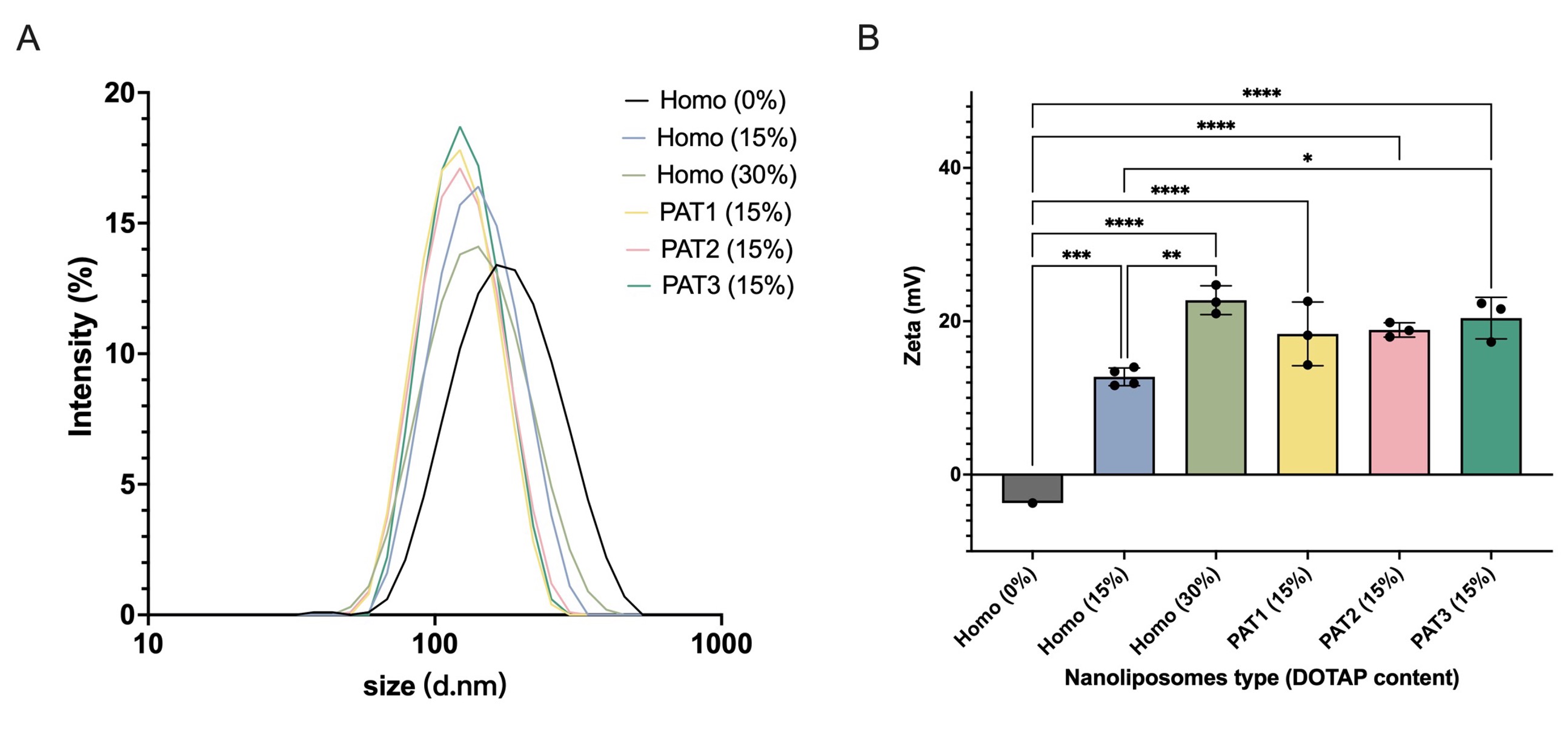}
	\caption{The size distribution (A) and zeta potential (B) of the examined SUVs of different lipid compositions. Percentage of DOTAP is presented in parenthesis. In bar graph, bars represent the average and error bars represent standard deviation with n$\geq$3. Data was statistically analyzed using one-way ANOVA and $^{****}$: p values $<$ 0.0001; $^{***}$: p values $<$ 0.001; $^{**}$: p values $<$ 0.01; $^{*}$: p value $<$ 0.05.
}
	\label{fig:nonpot}
\end{figure}

\vskip 1cm 
%\begin{table}[htb]
%\begin{center}
% \begin{tabular}{ | c |  c |  c |  c |  c |  c |}
%\hline
%Homo (0\%)  & Homo (15\%) & Homo (30\%) & PAT1 (15\%) & PAT2 (15\%)  & PAT3 (15\%)  \\
%\hline
%0 & 41.94 & 76.92 & 46.15 & 57.58 & 83.87\\
%%\hline
%0 & 36.67 & 100 & 50 & 74.07 & 81.25\\
%%\hline
% & 41.765 & 63.33 & 50 & 60.71 & 100\\
%%\hline
% & 68 & 96.67 &64.29 & 56 & 100 \\
%%\hline 
%& & 86.84 & & & \\
%\hline
%\end{tabular}
%\caption{Fusion data (\%) corresponding to Fig.~\ref{fig:fusion_frac}.
%}\label{tab:fusion_data}
%\end{center}
%\end{table}

\end{document}